# Two-dimensional magnetic nanoelectromechanical resonators

Shengwei Jiang[1], Hongchao Xie[1], Jie Shan[1,2,3*], Kin Fai Mak[1,2,3*]

[1]Laboratory of Atomic and Solid State Physics, Cornell University, Ithaca, NY, USA.
[2]School of Applied and Engineering Physics, Cornell University, Ithaca, NY, USA
[3]Kavli Institute at Cornell for Nanoscale Science, Ithaca, NY, USA
Email: jie.shan@cornell.edu; kinfai.mak@cornell.edu
These authors contributed equally: Shengwei Jiang, Hongchao Xie.

**Two-dimensional (2D) layered materials possess outstanding mechanical, electronic and optical properties, making them ideal materials for nanoelectromechanical applications [1,2]. The recent discovery of 2D magnetic materials [3,4,5,6,7,8] has promised a new class of magnetically active nanoelectromechanical systems (NEMS). Here we demonstrate resonators made of 2D $CrI_3$, whose mechanical resonances depend on the magnetic state of the material. We quantify the underlining effects of exchange and anisotropy magnetostriction by measuring the field dependence of the resonance frequency under a magnetic field parallel and perpendicular to the easy axis, respectively. Furthermore, we show efficient strain tuning of magnetism in 2D $CrI_3$ as a result of the inverse magnetostrictive effect using the NEMS platform. Our results establish the basis for mechanical detection of magnetism and magnetic phase transitions in 2D layered magnetic materials. The new magnetic NEMS may also find applications in magnetic actuation and sensing.**

Two-dimensional (2D) layered magnetic materials are attractive building blocks for nanoelectromechanical systems (NEMS): while they share high stiffness and strength and low mass with other 2D materials, they are magnetically active [1,2,3,6,7,8]. Among the large class of newly emerged 2D layered magnetic materials, of particular interest is few-layer $CrI_3$, whose magnetic ground state consists of antiferromagnetically coupled ferromagnetic (FM) monolayers with out-of-plane easy axis [3,5]. The interlayer exchange interaction is relatively weak, a magnetic field on the order of 0.5 T in the out-of-plane ($\hat{z}$) direction can induce spin-flip transition in bilayer $CrI_3$. Remarkable phenomena and device concepts based on detecting and controlling the interlayer magnetic state have been recently demonstrated, including spin-filter giant magnetoresistance [9,10,11,12], magnetic switching by electric field [13] or electrostatic doping [14,15,16,17], and spin transistors [18]. The coupling between the magnetic and mechanical properties in atomically thin materials, the basis for 2D magnetic NEMS, however, remains elusive although NEMS made of thicker magnetic materials [19] or coated with FM metals [20,21] have been studied.

Here we demonstrate magnetic NEMS resonators made of 2D $CrI_3$ drumhead membranes and investigate magnetostriction and its inverse effect in the material (Fig. 1a). We focus on even-layer CrI$_3$ for maximum effect. We encapsulate CrI$_3$ (an air sensitive insulator) within two stable 2D materials, few-layer graphene (below) and monolayer WSe$_2$ (above) (Fig. 1b). In addition to protect CrI$_3$ from degradation under ambient conditions, few-layer graphene acts as a conducting electrode, while monolayer WSe$_2$ as a strain gauge as we discuss below. The 2D heterostructure was first assembled and then transferred on prefabricated circular microtrenches (2 – 3 μm in radius) with patterned Au electrodes and Si back gate. The optical image of a sample device is shown in Fig. 1c. We drive mechanical resonances in the membrane electrically in the linear regime using a small r.f. gate voltage, which is superposed on a DC gate voltage $V_g$ that applies static tension to the membrane by pulling it towards the gate. We detect the mechanical resonances interferometrically at 633 nm using a network analyzer. We also measure the out-of-plane magnetization of the membrane using magnetic circular dichroism (MCD) at 633 nm. Unless otherwise specified, all measurements were performed at 4 K in a helium exchange gas environment (<10$^{-5}$ Torr). (See Methods for details on device fabrication and measurement techniques.)

Figure 1d shows the fundamental resonance mode of a bilayer CrI$_3$ membrane at $V_g$ = 0. It has a Lorentzian lineshape (solid line) with a peak frequency $f$ around 25.5 MHz and a width around 10 kHz, corresponding to a quality factor of about 2500. The resonance frequency scales inversely with the drumhead radius $R$, and is determined by stress on the membrane $\sigma$ and its 2D mass density $\rho$. Figure 1e shows gate tuning of the resonance frequency, which is well described by the continuum model (assuming zero bending stiffness) with fully clamped boundary: $f = \frac{\xi}{2\pi R}\sqrt{\frac{\sigma}{\rho}}$ (red solid line). Here $\xi \approx$ 2.405 is the first root of the zeroth order Bessel function and the stress is obtained by the second derivative of the total energy (elastic plus electrostatic) of the membrane with respect to its vertical displacement. Near $V_g$ = 0, the frequency is set by the built-in stress $\sigma_0$, whereas at higher $V_g$ the gate-induced stress $3Y_{eff}\epsilon(V_g)$, which is determined by the effective 2D Young's modulus $Y_{eff}$ and the gate-induced strain $\epsilon(V_g)$, causes the frequency to increase roughly as $V_g^4$. This behavior is fully consistent with other 2D NEMS such as graphene and MoS$_2$ [Ref. [22,23,24]]. By calibrating $\epsilon(V_g)$ from the exciton peak shift of the WSe$_2$ layer (63 meV per % of biaxial strain [25]), we determine the basic parameters of the resonator $\sigma_0 \approx$ 0.5 N/m, $Y_{eff} \approx$ 600 N/m and $\rho \approx 3 \times 10^{-5}$ kg/m$^2$ [Ref. [22,23,24]] (see Methods).

We investigate the effect of magnetic field on CrI$_3$ resonators at $V_g$ = 0. The nanomechanical resonance of the bilayer CrI$_3$ device is measured while the field is swept from 1 T to -1 T back to 1 T along $\hat{\mathbf{z}}$ (Fig. 2a). The resonance frequency $f$ is independent of field except an abrupt redshift of ~ 0.06% when the field magnitude exceeds ~ 0.5 T. The linewidth and amplitude basically do not change throughout the entire field range (Supplementary Fig. S1). We correlate the field dependence of $f$ (Fig. 2b) with the membrane MCD (Fig. 2c). The measured MCD is consistent with the reported results for bilayer CrI$_3$ [Ref. [14]]. It shows that the membrane is antiferromagnetic (AF) under small fields (the finite MCD signal is caused by layer asymmetry in the heterostructure) and undergoes a first-order AF-FM spin-flip transition (with hysteresis) around ±0.5 T. The mechanical resonance frequency is thus correlated with the sample's magnetic state with $f_{AF} > f_{FM}$, where $f_{AF}$ and $f_{FM}$ denote the resonance frequency in the AF and FM state, respectively. In contrast to MCD, $f$ is insensitive to the magnetization direction.

Similar behavior is observed in all CrI$_3$ resonators that were investigated in this study. Figure 2d - 2f shows the corresponding result for a 6-layer CrI$_3$ resonator under an out-of-plane field up to 2.3 T. A second 6-layer device is shown in Supplementary Fig. S2. Compared to bilayer CrI$_3$, there are now two spin-flip transitions around 0.9 T and 1.8 T, which correspond to spin flip in the surface layer and the interior layers, respectively [9,18]. The resonance frequency redshifts at each spin-flip transition. The total redshift from $f_{AF}$ (~ 0 T) to $f_{FM}$ (>1.8 T) is about 0.23%. This is nearly 5 times larger than in bilayer CrI$_3$.

In contrast, the behavior of the resonators under an in-plane magnetic field is distinct. Figure 3a shows the nanomechanical resonance of the same 6-layer CrI$_3$ device as in Fig. 2d - 2f while the magnetic field is swept from 8 T to -8 T to 8 T along the in-plane direction. The field dependence of $f$ is shown in Fig. 3b after subtraction of a small linear drift due to initial stress relaxation and/or slow temperature drift. Sharp transition and hysteresis that are characteristic for the first-order spin-flip transition are now absent. Instead, $f$ redshifts smoothly with increasing field magnitude and saturates beyond ~ 6 T. This behavior is correlated with the spin-canting effect observed in 2D CrI$_3$ [Ref. [9,18]]. Because of the large out-of-plane magnetic anisotropy, the spins under an in-plane field are canted continuously from $\hat{z}$ until reaching the saturation field, beyond which they are fully aligned with the in-plane field. The total change in $f$ under the in-plane field (~ 0.13%) is about half of the value under the out-of-plane field for the 6-layer CrI$_3$ resonator.

The resonance frequency of the NEMS at $V_g$ = 0 is determined by the built-in stress $\sigma_0$. The observed correlation between the resonance frequency and the magnetic state of 2D CrI$_3$ suggests that magnetostriction is a result of competition between minimizing the elastic energy and the internal magnetic interactions. Other effects such as arisen from the magnetostatic energy cannot explain the experimental observations (see Methods). The elastic energy of the membrane per unit area can be expressed as $U = \frac{3}{2} Y_{eff} \epsilon^2$ [Ref. [22,23,24]]. Here the effective 2D Young's modulus $Y_{eff}$ is dominated by that of few-layer graphene because of its much higher mechanical stiffness and is independent of magnetic field (Methods); strain $\epsilon = (a - a_0)/a_0$ is defined as the fractional change of the in-plane lattice constant $a$ that conforms to fixed boundary from its equilibrium value $a_0$. If we neglect the strain dependence of the demagnetization energy and leave out the intralayer exchange interactions, which do not play a role in the interlayer metamagnetic transition, the part of free energy that is strain dependent can be expressed for bilayer CrI$_3$ to the lowest relevant order as

$$F = J_\perp(\widehat{\boldsymbol{S}}_t \cdot \widehat{\boldsymbol{S}}_b) + K_{eff}\left(|\widehat{\boldsymbol{S}}_t \times \hat{\boldsymbol{z}}|^2 + |\widehat{\boldsymbol{S}}_b \times \hat{\boldsymbol{z}}|^2\right). \tag{1}$$

Here $\widehat{\boldsymbol{S}}_t$ and $\widehat{\boldsymbol{S}}_b$ denote the spin unit vector of the top and bottom CrI$_3$ layers, respectively. The two terms describe, respectively, the interlayer exchange coupling with energy per unit area $J_\perp$ (> 0) and anisotropy with effective energy per unit area per layer $K_{eff}$ (> 0). For simplicity we only consider the effect of bond length change on $J_\perp$ and $K_{eff}$. The equilibrium lattice constant of the strained membrane in each magnetic state can be found by minimizing the total free energy ($U + F$) with respect to $a_0$ or equivalently strain $\epsilon$, from which stress ($\sigma_0 = Y_{eff}\epsilon$) and mechanical resonance frequency can be evaluated [26] (see Methods).

For spin-flip transitions, the anisotropy energy is not relevant because spins are along the easy axis. Minimizing ($U + F$) for bilayer CrI$_3$ yields a change in the strain level ($-\frac{2}{3Y_{eff}}\frac{\partial J_\perp}{\partial \epsilon}$) and a fractional change in the resonance frequency ($-\frac{1}{3\sigma_0}\frac{\partial J_\perp}{\partial \epsilon}$) for the AF-FM transition. We estimate $\frac{\partial J_\perp}{\partial \epsilon}A_u \approx -3$ meV, where $A_u$ ($\approx 0.47$ nm$^2$) is the unit cell area, from the experimental resonance frequency shift and the measured built-in stress $\sigma_0$. The negative sign indicates that strain weakens AF ordering. We also estimate the saturation magnetostriction $|\frac{2}{3Y_{eff}}\frac{\partial J_\perp}{\partial \epsilon}| \sim 10^{-6}$ for the bilayer CrI$_3$ heterostructure. We note that bilayer CrI$_3$ by itself is expected to have a much higher saturation magnetostriction because of its much smaller Young's modulus compared to graphene.

In 6-layer CrI$_3$ there are 5 interfaces. The nearly 5-time stronger magnetostrictive response observed in experiment can be understood in terms of a 5-time stronger total interlayer exchange energy $5J_\perp(\widehat{\boldsymbol{S}}_t \cdot \widehat{\boldsymbol{S}}_b)$ and similar elastic energy compared to the bilayer CrI$_3$ device. Under an in-plane field, both interlayer exchange and anisotropy contributions to the free energy need to be considered and a similar energy minimization scheme can be applied to yield $-\frac{1}{3\sigma_0}\left(5\frac{\partial J_\perp}{\partial \epsilon} + 6\frac{\partial K_{eff}}{\partial \epsilon}\right)$ for the fractional resonance frequency shift. The observed ~ 50% smaller frequency shift in Fig. 3b compared to Fig. 2e suggests that $\frac{\partial K_{eff}}{\partial \epsilon} \approx -\frac{1}{2}\frac{\partial J_\perp}{\partial \epsilon}$ in the 6-layer CrI$_3$ device.

Finally, we use the NEMS platform to demonstrate strain tuning of magnetism through the inverse magnetostrictive effect. We probe the spin-flip transition in the same bilayer CrI$_3$ device as in Fig. 2a-c at different $V_g$'s by MCD. The application of $V_g$ can potentially induce both tension by an electrostatic force and electrostatic doping to the suspended membrane. Figure 4a and 4b compare the behavior of a suspended and substrate-supported region of the membrane. Only the MCD data near the spin-flip transition for the positive sweeping direction of the field is shown for clarity. For the substrate-supported region, only the electrostatic doping effect is relevant. The spin-flip transition field $H_C$ varies linearly with $V_g$ (red symbols, Fig. 4c), which is in good agreement with previous studies [14,15]. In contrast, the behavior of the suspended region is symmetric about $V_g = 0$; $H_C$ decreases nonlinearly with magnitude of $V_g$ (blue symbols, Fig. 4c). We conclude from the comparison that electrostatic doping into suspended CrI$_3$ is negligible and the change in $H_C$ is caused primarily by strain.

Figure 4d shows $H_C$ as a function of gate-induced strain calibrated from the exciton peak shift in the WSe$_2$ layer. It decreases linearly with strain by as much as ~ 32 mT. Quantitatively, $H_C$ can be related to the interlayer exchange coupling $J_\perp$ through $\mu_0 H_C = \frac{J_\perp + \mu_0 M_0^2/d}{M_0}$ [Ref. 14], where $\mu_0$, $M_0$ and $d$ denote the vacuum permeability, the saturation magnetization per CrI$_3$ monolayer, and the interlayer distance in bilayer CrI$_3$, respectively (see Methods). We obtain $\frac{\partial J_\perp}{\partial \epsilon} A_u \approx -5$ meV from the slope in Fig. 4d and $M_0$ by assuming that each Cr$^{3+}$ cation carries a magnetic moment of $3\mu_B$ (Bohr magneton). The value agrees well with that estimated from the mechanical resonance measurement. The discrepancy is largely due to uncertainty in the resonator parameters (such as initial stress and mass). Nevertheless, the good agreement between two independent measurements illustrates the importance of interlayer exchange magnetostriction in 2D CrI$_3$.

In conclusion, we have demonstrated a new type of magnetostrictive NEMS based on 2D CrI$_3$. An inverse magnetostrictive effect has also been demonstrated, which allows continuous strain tuning of the internal magnetic interactions. Our results have put in place the groundwork for potential applications of these devices, including magnetic actuation and sensing, as well as a general detection scheme based on mechanical resonances for emerging magnetic states and phase transitions in 2D layered magnetic materials.

**Methods**
**Device fabrication**
We used the layer-by-layer dry transfer method to fabricate drumhead resonators made of atomically thin CrI$_3$ fully encapsulated by few-layer graphene and single-layer WSe$_2$ as shown in Fig. 1b. Atomically thin flakes of CrI$_3$, graphene and WSe$_2$ were mechanically exfoliated onto silicon substrates with a 285-nm oxide layer from the corresponding bulk synthetic crystals (from HQ Graphene). A polymer stamp made of a thin layer of polycarbonate (PC) on polydimethylsiloxane (PDMS) was used to pick up the desired flakes one by one to form the heterostructure. The complete heterostructure was first released onto a new PDMS substrate so that the residual PC film on the sample can be removed by dissolving it in N-Methyl-2-pyrrolidone. The sample was then transferred onto a circular microtrench of 2 – 3 $\mu$m in radius and 600 nm in depth on a silicon substrate with prepatterned Ti/Au electrodes. A small amount of PDMS residual was left untreated. CrI$_3$ was handled inside a nitrogen-filled glovebox with oxygen and water less than 1 part per million (ppm) to avoid degradation. The thickness of atomically thin materials was first estimated by their optical reflection contrast and then measured by atomic force microscopy (AFM). The thickness of CrI$_3$ flakes was further verified by the magnetization versus out-of-plane magnetic field measurement. The thickness of WSe$_2$ flakes was verified by optical reflection spectroscopy.

**Mechanical resonance detection**
An optical interferometric technique was applied to detect the out-of-plane displacement of the resonators [22,24,27] (Fig. 1a). The resonators were mounted in a closed-cycle cryostat (Attocube, attoDry1000). The output of a HeNe laser at 632.8 nm was focused onto the center of the suspended membrane using a high numerical aperture (N.A. = 0.8) objective. The beam size on the device was on the order of 1 $\mu$m and the total power was kept below 1 $\mu$W to minimize the laser heating effect. The reflected beam was collected by the same objective and detected by a fast photodetector. Motion of the membrane was actuated capacitively by applying a small r.f. gate voltage (~ 1 mV) between the membrane and the back gate. As the membrane moves in the out-of-plane direction, the optical cavity formed between the membrane and the trenched substrate modulates the device reflectance. The amplitude of the motion as a function of driving frequency was measured by a network analyzer (Agilent E5061A), which both provided the r.f. voltage and measured the fast photodetector response. The amplitude of the motion reaches its maximum when the r.f. frequency matches the natural frequency of the resonator.

**Magnetic circular dichroism (MCD) measurements**
MCD was employed to characterize the membranes' magnetic properties. A nearly identical optical setup as the one used for the mechanical resonance detection was employed. The polarization of the optical excitation at 632.8 nm was modulated between left and right circular polarization by a photoelastic modulator at 50.1 kHz. Both the a.c. and d.c. component of the reflected beam were detected by a photodiode in combination with a lock-in amplifier and multimeter, respectively. MCD was determined as the ratio of the a.c. and d.c. component.

**Strain calibration**
Gate-induced strain in the suspended membranes was determined by using monolayer WSe$_2$ as a strain gauge assuming no relative sliding between the constituent 2D layers. It relies on the fact that the fundamental exciton resonance energy in monolayer WSe$_2$ redshifts linearly with strain (63 meV/%) for a relatively large rage of biaxial strain [25]. The fundamental exciton energy in monolayer WSe$_2$ was determined as a function of $V_g$ by optical reflection spectroscopy. In these measurements, broadband radiation from a single-mode fiber-coupled halogen lamp was employed. The collected radiation was detected by a spectrometer equipped with a charge-coupled-device (CCD) camera. The excitation power on the device was kept well below 0.1 $\mu$W to minimize the laser heating effect. The exciton resonance energy and the calibrated gate-induced strain as a function of $V_g$ is shown in Supplementary Fig. S3.

**Characterization of resonator parameters**

By minimizing the sum of the elastic energy and the electrostatic energy with respect to strain (i.e. $\frac{\partial}{\partial \epsilon}\left[\frac{3}{2}Y_{eff}\epsilon^2 - \frac{1}{2}C_g V_g^2\right] = 0$), we obtain the gate-induced strain $\epsilon(V_g) \approx \frac{1}{96}\left(\frac{R}{D}\right)^2 \left(\frac{\epsilon_0 V_g^2}{D\sigma_0}\right)^2 \propto V_g^4$ for $\epsilon(V_g) \ll \epsilon_0$ ($\epsilon_0$ is the built-in strain). Here $C_g$ is the back gate capacitance, which is strain or gate dependent because of the gate-induced vertical displacement of the membrane, $D$ is the vertical separation between the membrane and the back gate at $V_g = 0$, $R$ is the drumhead radius, and $\varepsilon_0$ is the vacuum permittivity. The built-in stress $\sigma_0$ can be obtained from the slope of $\epsilon$ as a function of $V_g^4$ (Supplementary Fig. S3c). The effective Young's modulus $Y_{eff}$ and the 2D mass density $\rho$ of the membrane can be obtained by fitting the experimental gate dependence of the resonance frequency $f = \frac{\xi}{2\pi R}\sqrt{\frac{\sigma}{\rho}}$ with $\sigma = \frac{R^2}{4}\frac{\partial^2}{\partial z^2}\left[\frac{3}{2}Y_{eff}\epsilon^2 - \frac{1}{2}C_g V_g^2\right]$, where $z$ is the vertical displacement [Ref. [28,29,30]]. The extracted values for bilayer CrI3 resonator 1 are $\sigma_0 \approx 0.5\ Nm^{-1}$, $\rho \approx 3 \times 10^{-5}\ kgm^{-2}$, and $Y_{eff} \approx 600\ Nm^{-1}$. The mass density is 1.9 times of the expected mass density of the heterostructure presumably due to the presence of polymer residues and other adsorbates on the membrane. The effective Young's modulus is also consistent with the reported values. We estimate the effective Young's modulus of 2D heterostructures as the total contribution of constitute layers, $Y_{eff} = \sum_i n_i Y_{2D,i}$, where $n_i$ and $Y_{2D,i}$ are the layer number and the 2D elastic stiffness of the $i$-th material per layer. We obtain $Y_{eff} \approx 1200\ Nm^{-1}$ for bilayer device 1 using $Y_{2D,g} = 340\ Nm^{-1}$, $Y_{2D,CrI_3} = 25\ Nm^{-1}$ and $Y_{2D,WSe_2} = 112\ Nm^{-1}$ reported for the constitute 2D materials [Ref. [31,32,33,34]]. The discrepancy may come from the presence of polymer residues and/or wrinkles on the membrane. Most devices studied in this work consist of 3 layers of graphene, 1 layer WSe2, and 2-6 layers of CrI3. $Y_{eff}$ is therefore dominated by the contribution of graphene.

**Mechanisms for mechanical resonance shift in 2D CrI3 under a magnetic field**

In the main text, we have assigned the competition between the internal magnetic interactions and elastic energy as the major mechanism for the observed mechanical resonance shift in 2D CrI3 resonators under a magnetic field at $V_g = 0$. Other effects could potentially also give rise to a mechanical resonance shift in 2D CrI3 under a magnetic field. One such possibility is the magnetostatic pressure from the gradient of the magnetostatic energy in the $\hat{z}$ direction, $\boldsymbol{M} \cdot \frac{\partial \boldsymbol{B}}{\partial z}$. Here $\boldsymbol{M}$ is the sample magnetization and $\boldsymbol{B}$ is the magnetic field at the sample. However, for a nanometer thick sample with a lateral size of a few microns the gradient is expected to be negligible. This is supported by the absence of field dependence for the mechanical resonance up to high field except at the spin-flip transitions (Supplementary Fig. S4).

**Model for exchange magnetostriction in bilayer CrI3**

We consider bilayer CrI3 under an out-of-plane magnetic field $\mu_0 H$. The free energy per unit area for a free membrane in the AF and FM state in the zero-temperature limit can be expressed as

$$F_{AF} = 2F_0 - J_\perp, \tag{S1}$$

$$F_{FM} = 2F_0 + J_\perp - 2\mu_0 M_0 \left(H - \frac{M_0}{t}\right). \tag{S2}$$

Here $F_0$, $M_0$, $J_\perp$ and $t$ denote the free energy of each monolayer, saturation magnetization of each monolayer, interlayer exchange interaction, and interlayer distance, respectively. The spin-flip field $\mu_0 H_C$ can be evaluated by requiring $F_{AF} = F_{FM}$ [Ref. [13]]. The equilibrium lattice constant in the two interlayer magnetic states differs slightly in free membranes. In the presence of fixed boundary, strain is developed with an elastic energy of $U = \frac{3}{2} Y_{eff} \epsilon^2$, where $\epsilon = (a - a_0)/a_0$ is the fractional change of the in-plane lattice constant $a$ (which is fixed by the boundary) from its equilibrium value $a_0$. A new equilibrium configuration with lattice constants $a_{AF}$ and $a_{FM}$ is reached for the two magnetic states to minimize the corresponding total free energy $(U + F)$ with respect to $a_0$. Strain in the AF and FM states $\epsilon_{AF}$ and $\epsilon_{FM}$ can thus be found by solving the following equations:

$$a_0 \frac{\partial (U + F_{AF})}{\partial a_0} = 2 \frac{\partial F_0}{\partial \epsilon} - \frac{\partial J_\perp}{\partial \epsilon} - 3 Y_{eff} \epsilon_{AF} = 0. \tag{S3}$$

$$a_0 \frac{\partial (U + F_{FM})}{\partial a_0} = 2 \frac{\partial F_0}{\partial \epsilon} + \frac{\partial J_\perp}{\partial \epsilon} - 3 Y_{eff} \epsilon_{FM} = 0. \tag{S4}$$

The derivatives are evaluated at the equilibrium lattice constants. Note that $a_0 \frac{\partial U}{\partial a_0} = -\frac{\partial U}{\partial \epsilon}$ for the fixed boundary condition (i.e. $a$ is a constant), and the strain dependence of $M_0$ is ignored. Therefore, the strain difference between the two states is

$$\epsilon_{AF} - \epsilon_{FM} = -\frac{2}{3 Y_{eff}} \frac{\partial J_\perp}{\partial \epsilon}. \tag{S5}$$

And the fractional change in the resonance frequency is given by

$$\frac{f_{AF} - f_{FM}}{f_{AF}} \approx \frac{\sigma_{AF} - \sigma_{FM}}{2\sigma_{AF}} = -\frac{1}{3\sigma_0} \frac{\partial J_\perp}{\partial \epsilon}, \tag{S6}$$

since $f \propto \sqrt{\sigma}$. Here we have taken the initial state to be AF. The effect of magnetic anisotropy under in-plane magnetic field can be taken into account in a similar manner.

**Competing interests**

The authors declare no competing interests.

**Data availability**

The data that support the findings of this study are available within the paper and its Supplementary Information. Additional data are available from the corresponding authors upon request.

**Figures and figure captions**

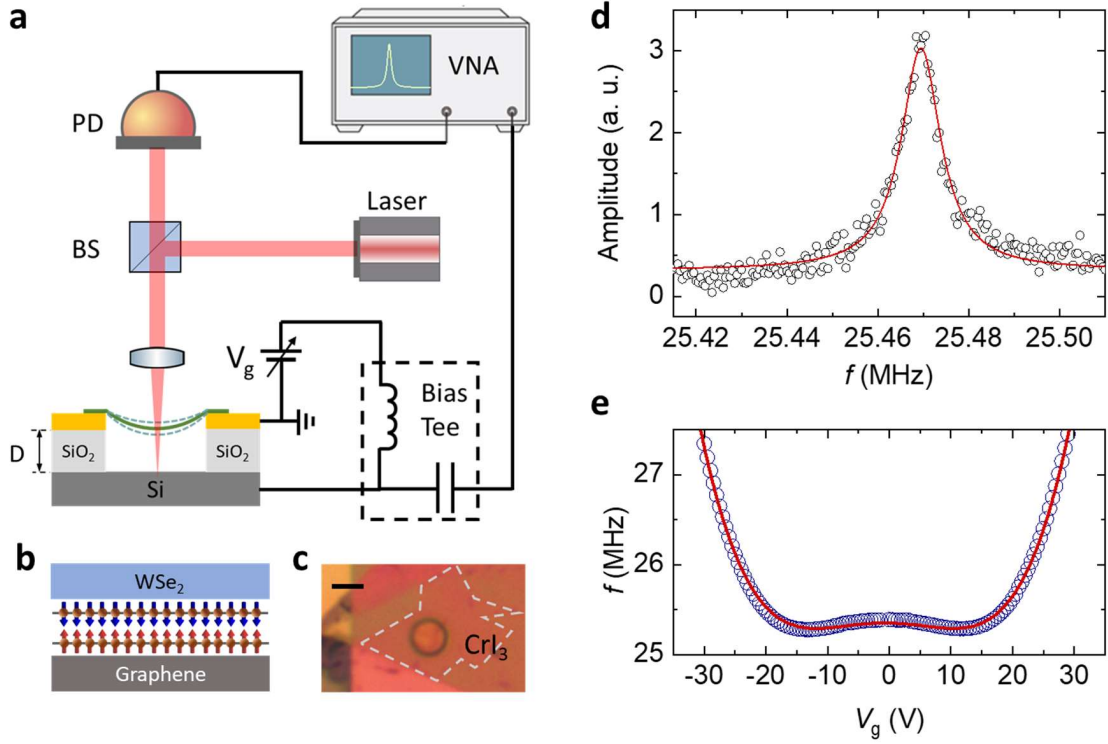

**Fig. 1: 2D CrI$_3$ nanoelectromechanical resonators. a,** Schematic of the measurement system. The resonator (suspended 2D membrane on a Si trench of depth *D*) is actuated by an r.f. voltage from a vector network analyzer (VNA) through a bias tee. A DC voltage $V_g$ is superimposed to apply static tension to the membrane. The motion is detected interferometrically by a HeNe laser, which is focused onto the center of the resonator. BS: beam splitter; PD: photodetector. **b,** Schematic of a bilayer CrI$_3$ resonator with AF CrI$_3$ encapsulated by few-layer graphene and monolayer WSe$_2$. Filled spheres and arrows denote Cr atoms and spins in the top and bottom CrI$_3$ layer. **c,** Optical microscope image of a bilayer CrI$_3$ device suspended over a circular trench. Dashed line shows the boundary of the CrI$_3$ flake. Scale bar is 4 μm. **d,** Fundamental mechanical resonance (symbols) of bilayer CrI$_3$ resonator 1 (radius 2 μm) and a Lorentzian fit of the resonance spectrum (solid line). **e,** Gate dependence of the measured resonance frequency (symbols) and fit to the continuum model (solid line) with $\sigma_0 \approx 0.5\ Nm^{-1}$, $\rho \approx 3 \times 10^{-5}\ kgm^{-2}$ (1.9 times the mass density of the membrane) and $Y_{eff} \approx 600\ Nm^{-1}$.

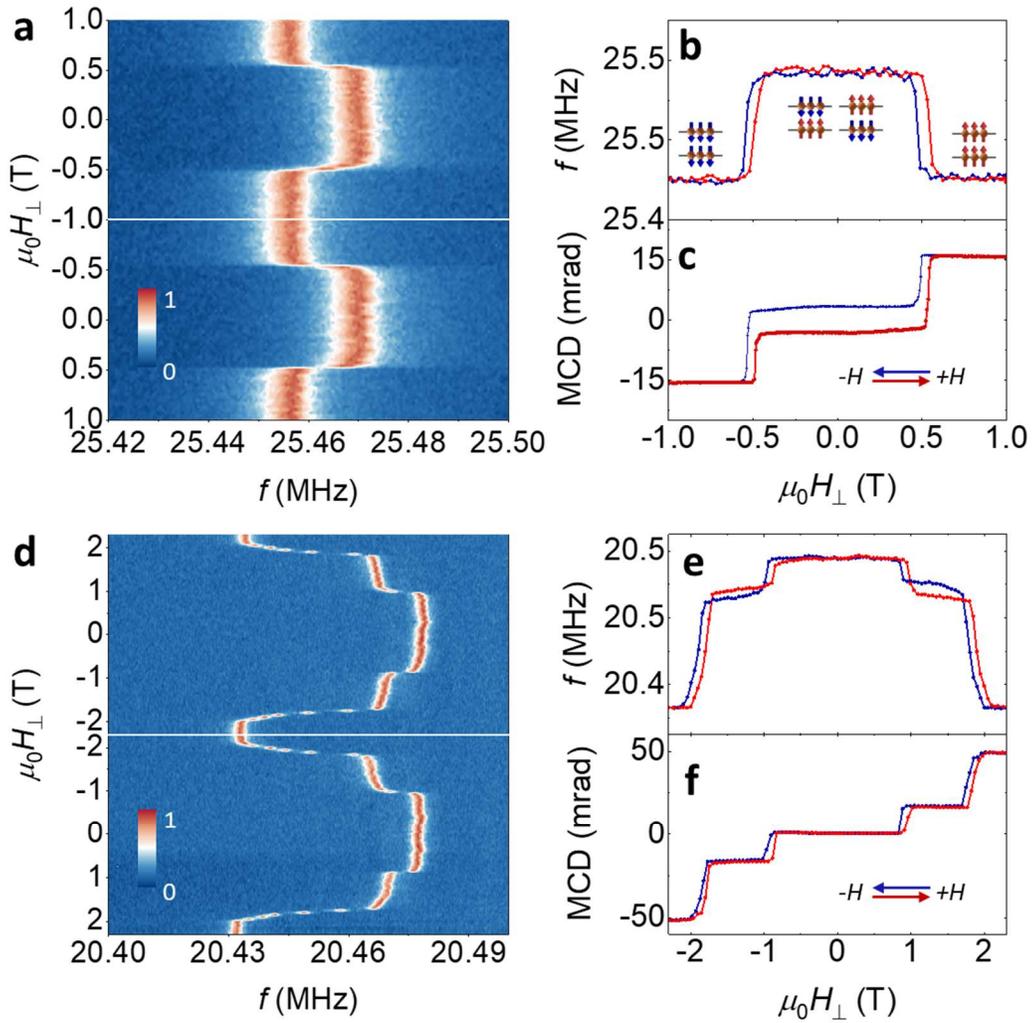

**Fig. 2: Mechanical detection of the spin-flip transition in 2D CrI$_3$. a**, Normalized vibration amplitude of bilayer CrI$_3$ resonator 1 vs. driving frequency under an out of-plane magnetic field ($\mu_0 H_\perp$) that sweeps from 1 T to -1 T to 1 T. **b**, Resonance frequency $f$ extracted from **a** as a function of magnetic field. **c**, MCD of the membrane as a function of magnetic field. The red and blue lines in **b** and **c** correspond to the measurement for the positive and negative sweeping directions of the field, respectively. **d-f**, Same measurements as in **a-c** for 6-layer CrI$_3$ resonator 1 (radius 3 μm).

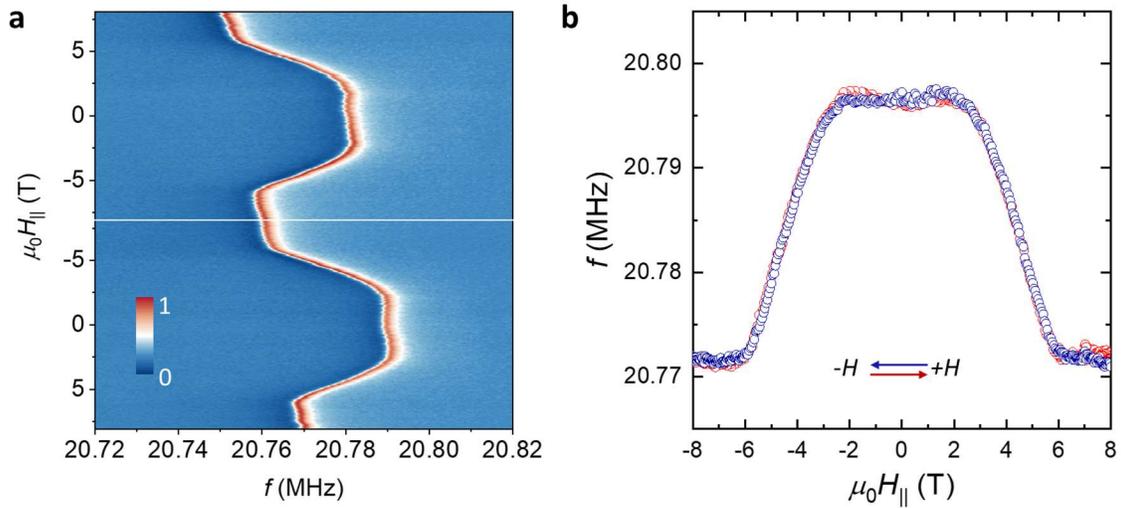

**Fig. 3: Mechanical detection of spin canting in 2D CrI$_3$. a**, Normalized vibration amplitude vs. driving frequency under an in-plane magnetic field ($\mu_0 H_\parallel$) that sweeps from 8 T to -8 T to 8 T for 6-layer CrI$_3$ resonator 1. **b**, Resonance frequency $f$ extracted from **a** as a function of magnetic field. Red and blue lines correspond to the measurement for the positive and negative sweeping directions of the field, respectively.

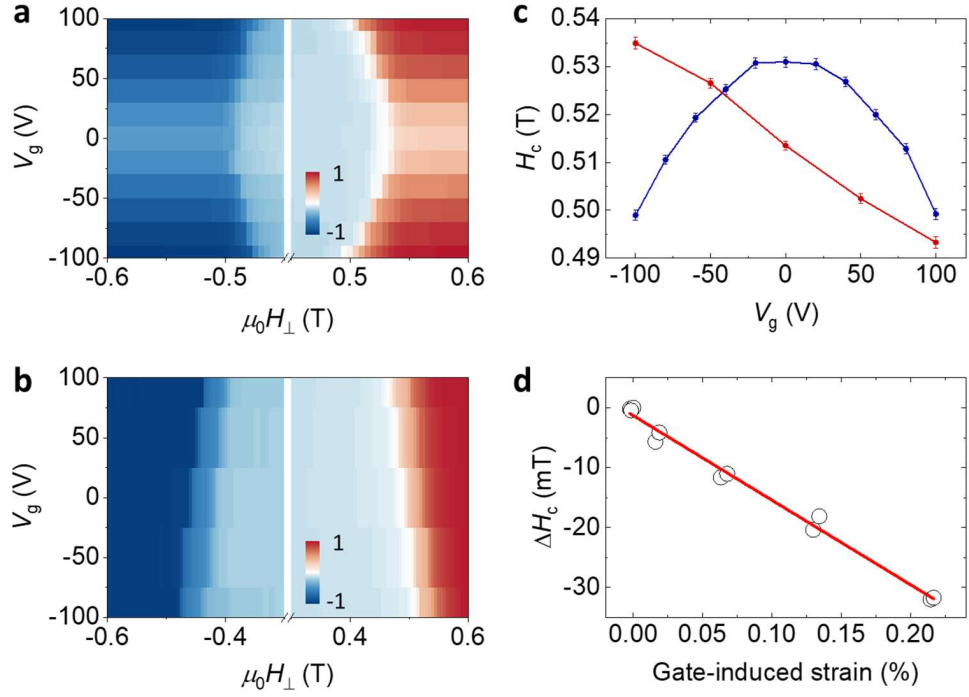

**Fig. 4: Strain tuning of the spin-flip transition in 2D CrI$_3$. a**, **b**, Normalized MCD as a function of out-of-plane magnetic field ($\mu_0 H_\perp$) that sweeps from - 1 T to 1 T (only - 0.6 T to 0.6 T is shown here for clarity) at different $V_g$'s for a suspended (**a**) and substrate-supported region (**b**) of bilayer CrI$_3$ resonator 1. **c**, Spin-flip transition field as a function of gate voltage for the suspended (blue) and substrate-supported (red) region of the membrane. The lines are a guide to the eye. **d**, Spin-flip transition field as a function of gate-induced strain (symbols). The solid line is a linear fit.

**Supplementary figures and figure captions**

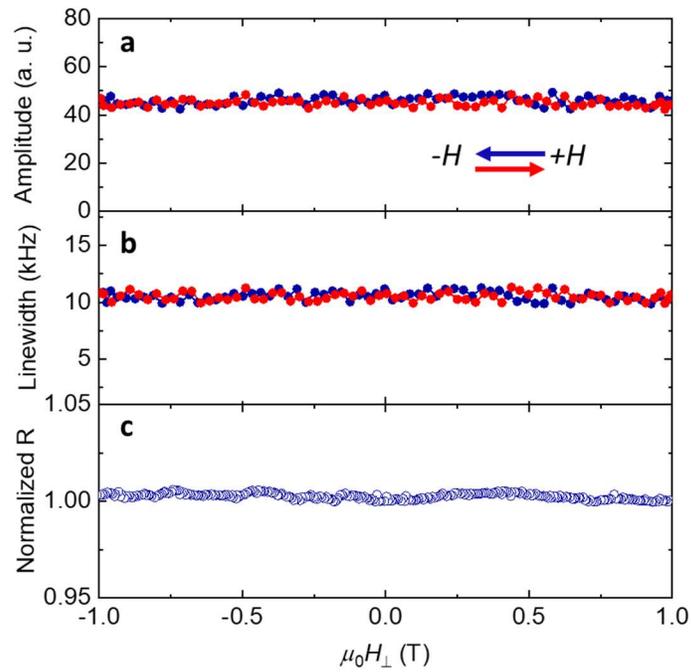

**Supplementary Fig. S1: Mechanical resonance of 2D CrI$_3$ under an out-of-plane magnetic field. a, b,** Field dependence of the amplitude (**a**) and linewidth (**b**) of the fundamental resonance of bilayer CrI$_3$ resonator 1. The field dependence of the resonance frequency is shown in Fig. 2b. The red and blue symbols correspond to the measurement for the positive and negative sweeping directions of the field, respectively. **c,** Field dependence of the reflectance at 633 nm normalized by the reflectance at zero field.

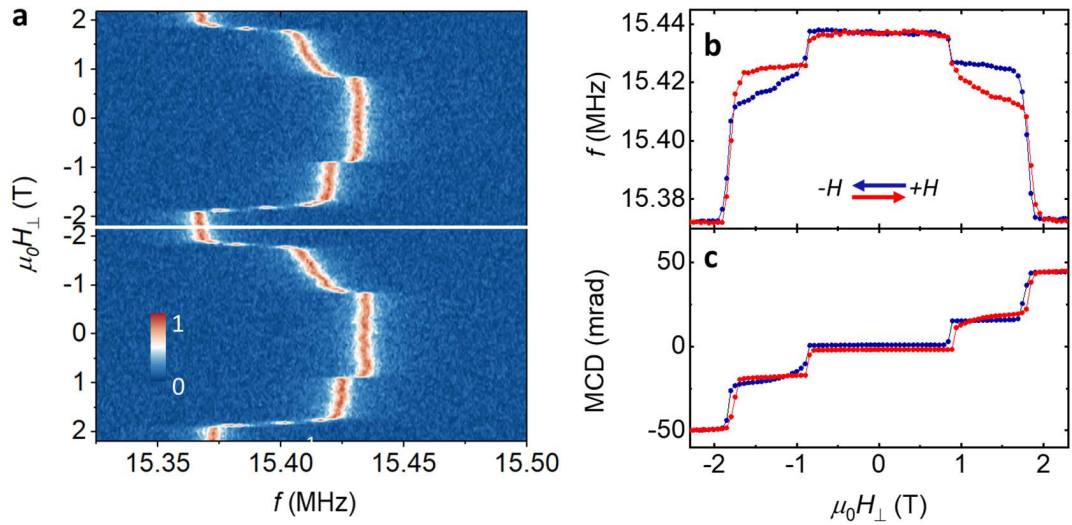

**Supplementary Fig. S2: Results for 6-layer CrI$_3$ resonator 2 under an out-of-plane magnetic field. a**, Normalized vibration amplitude vs. driving frequency under an out of-plane magnetic field ($\mu_0H_\perp$) that sweeps from 2.3 T to -2.3 T to 2.3 T at $V_g = 0$ V. **b**, Resonance frequency $f$ extracted from **a** as a function of magnetic field. **c**, MCD of the membrane as a function of magnetic field at $V_g = 0\,V$. The red and blue lines in **b, c** correspond to the measurement for the positive and negative sweeping directions of the field, respectively. The radius of the drumhead is 3 μm.

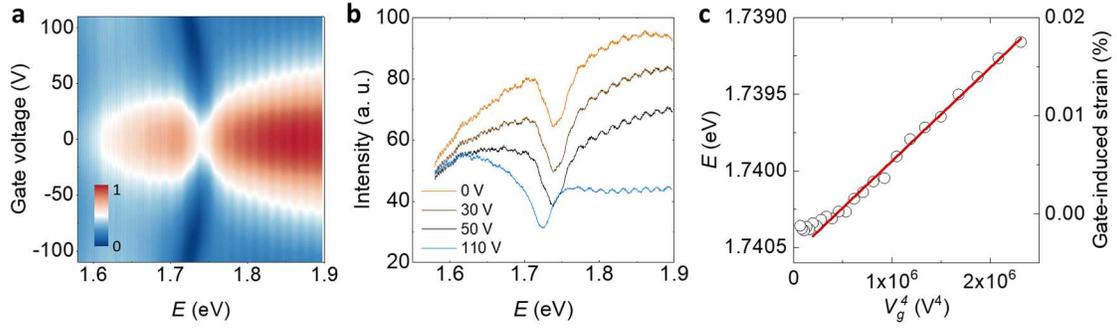

**Supplementary Fig. S3: Strain calibration and determination of resonator parameters. a**, Reflection contrast of bilayer CrI$_3$ resonator 1 from 1.6 – 1.9 eV as a function of gate voltage. The main feature is a dip (around 1.75 eV at $V_g = 0\,V$), which corresponds to the fundamental exciton resonance of monolayer WSe$_2$. The feature redshifts slightly with $V_g$ up to about 40 V followed by a much larger redshift with further increase of $V_g$. **b**, Representative spectra at selected $V_g$. **c**, Exciton resonance energy extracted from **a** (left axis) and gate-induced strain calibrated from the exciton resonance energy (right axis) as a function of $V_g^4$ for $V_g$ up to 39 V. The solid line is a linear fit from the slope of which the built-in stress $\sigma_0$ was determined.

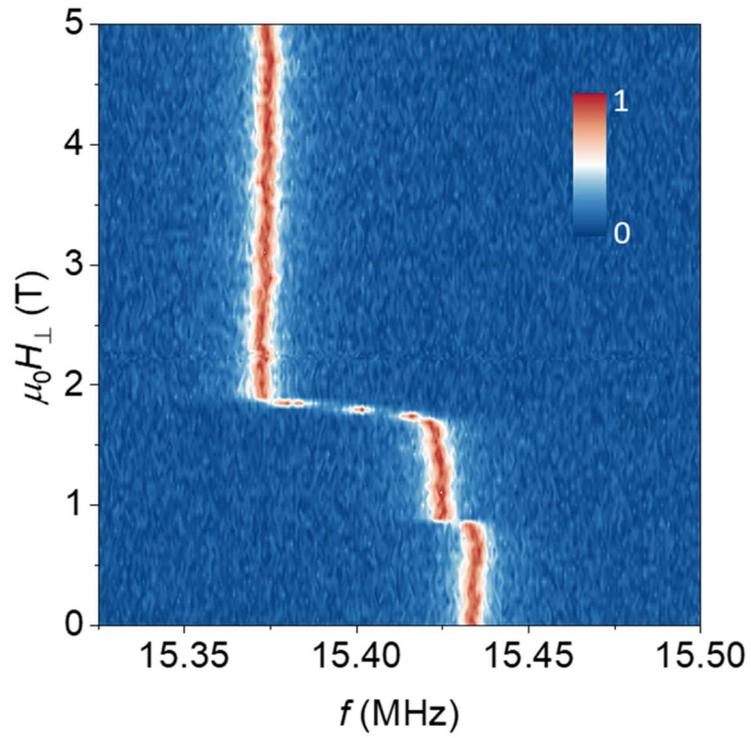

**Supplementary Fig. S4: 2D CrI₃ resonators under high out-of-plane fields.** Normalized vibration amplitude vs. driving frequency under an out of-plane magnetic field up to 5 T for 6-layer CrI$_3$ resonator 1. The resonance frequency, amplitude and linewidth basically do not change up to 5 T except at the spin-flip transitions.